\documentstyle[fleqn,11pt]{article}
\begin{document}

\begin{titlepage}
\vspace*{1.9cm}

\begin{center}
{\LARGE   \bf  $U(1)$ Connection, Nonlinear Dirac-like Equations
and Seiberg-Witten Equations}
\end{center}
\vspace*{0.5cm}

\begin{center}
{\bf  Liangzhong Hu\footnote{Institute of Mathematics, Peking University,
          Beijing 100871, People's Republic of China. \\
  E-mail address: lhu@sxx0.math.pku.edu.cn}
 and Liangyou Hu\footnote{Basic Department, Dongbei University of Finance
 and Economics, Dalian 116025, People's Republic of China.}
           }\\
\end{center}
\vspace*{0.5cm}
\begin{abstract}
By analysing the work of Campolattaro we argue that the second Seiberg-Witten
 equation over the Spin$^c_4$ manifold, i.e.,
 $F^+_{ij}=\langle M,S_{ij} M \rangle$, is the generalization of the
 Campolattaro's description of the electromagnetic field tensor $F^{\mu\nu}$
in the bilinear form $F^{\mu\nu}=\overline{\Psi} S^{\mu\nu}\Psi$.
 It turns out that the Seiberg-Witten equations (also the perturbed Seiberg-Witten
 equations) can be well understood from
 this point of view. We suggest that the second Seiberg-Witten equation
  can be replaced by a nonlinear Dirac-like Equation. We also derive the
  spinor representation of the connection on the associated
 unitary line bundle over the Spin$^c_4$ manifold.
\end{abstract}


\end{titlepage}

\section{INTRODUCTION}

The Maxwell equations and the Dirac equation are among the most celebrated
equations of physics. Campolattaro (1980a, 1980b) started with
the analysis of the Maxwell
equations by writing the electromagnetic field tensor $F_{\mu\nu}$ in
the equivalent bilinear form
\begin{equation}
F^{\mu\nu}=\overline{\Psi} S^{\mu\nu}\Psi
\end{equation}
where $\mu, \nu  = 0,  1,  2, 3.$
$\Psi$ is a Dirac spinor, $\overline{\Psi}=\Psi^{\dagger}\gamma^{0}$ is
the Dirac conjugation of $\Psi$. $S^{\mu\nu}$ is the spin operater defined
by
\begin{equation}
S^{\mu\nu} = \frac{i}{4}[\gamma^{\mu},\gamma^{\nu}]
\end{equation}
 and the $\gamma$'s are the Dirac matrices satisfying
\begin{equation}
     \gamma^{\mu} \gamma^{\nu} + \gamma^{\nu} \gamma^{\mu} = 2 \eta^{\mu \nu}
\end{equation}
with $\eta^{\mu\nu}$ the Minkowski metric tensor given by $\eta^{\mu\nu}
= diag (1, -1, -1, -1).$

In this representation the dual tensor
\begin{equation}
\ast F^{\mu\nu}=\overline{\Psi} \gamma^{5} S^{\mu\nu}\Psi
\end{equation}
From now  on, the Einstein sum convention is adopted throughout.
The Maxwell equations read (a comma followed by an index represents the partial
derivative with respect to the variable with that index)
\begin{eqnarray}
    (\overline{\Psi} S^{\mu\nu}\Psi)_{,\mu} &=& j^{\nu}     \\
    (\overline{\Psi} \gamma^{5} S^{\mu\nu}\Psi)_{,\mu} &=& 0
\end{eqnarray}
Moreover, the duality (Rainich, 1925; Misner and Wheeler, 1957) by the
complexion $\alpha$, namely
\begin{equation}
\overline{F}^{\mu\nu} = F^{\mu\nu} \cos{\alpha} + \ast F^{\mu\nu} \sin{\alpha}
\end{equation}
is equivalent to a Touschek-Nishijima (Touschek, 1957; Nishijima, 1957)
transformation for the spinor $\Psi$ to the spinor $\Psi'$ given by
\begin{equation}
         \Psi'= e^{\gamma^{5}\alpha/2}\Psi
\end{equation}
with
\begin{equation}
e^{\gamma^{5}\alpha}=\cos\alpha+\gamma^{5}\sin\alpha
\end{equation}
and
\begin{eqnarray}
 \cos\alpha &=& \frac{\overline{\Psi}\Psi}{\rho}       \\
 \sin\alpha &=& \frac{\overline{\Psi}\gamma^{5}\Psi}{\rho}
\end{eqnarray}
$\rho$ being the positive square root of
\begin{equation}
 \rho^{2} = (\overline{\Psi}\Psi)^{2} + (\overline{\Psi}\gamma^{5}\Psi)^{2}
\end{equation}
Campolattaro showed that the two spinor Maxwell equations (5) and (6) is
equivalent to a
single nonlinear first-order equation for the spinor, namely
\begin{equation}
 \gamma^{\mu}\Psi_{,\mu} = -i\gamma^{\mu}\frac{e^{\gamma^{5}\alpha}}{\rho}
  \{ Im(\overline{\Psi}_{,\mu}\Psi)- j_{\mu}-\gamma^{5}Im(\overline{\Psi}_{,\mu}
  \gamma^{5}\Psi)\} \Psi
\end{equation}
The relation between Dirac and Maxwell equations was also considered by Vaz
and Rodrigues (1993).

On the other hand, Witten (1994) introduced the Seiberg-Witten equations.
For more details, see, e. g., (Moore, 1996; Morgan, 1996).
By counting the solutions of the equations with an Abelian
 gauge group, new invariants of 4-manifolds can be obtained. These invariants
are closely related to Donaldson's polynomial invariants, but in many respects
much simpler to work with.
One of the Seiberg-Witten equation is the Dirac equation on the
Spin$^c_4$ manifold $X$,
\begin{equation}
 \widetilde{D}M = \gamma_{i} \widetilde{\nabla_{i}} M = 0
\end{equation}
where the Riemannian indices $i = 1, 2, 3, 4$.
      $\widetilde{D}$ is the Dirac operator on the Spin$^c_4$ manifold $X$.
      $ M \in S^+$, $S^+$ is the positive chirality spinor bundle
       over $X$.
$ \widetilde{\nabla_{i}} $ is a covariant derivative acting on $S^{+}$.
        $\gamma_{i}$, the Clifford matrices satisfy
\begin{equation}
       \gamma_{i} \gamma_{j}+\gamma_j \gamma_i = - 2 \delta_{i j}
\end{equation}
 Locally,
\begin{eqnarray}
      S^{+}&=&S^{+}_{0}\otimes L^{\frac{1}{2}}                 \\
      \widetilde{\nabla_{i}}
      &=&\nabla_{i}\otimes1+1\otimes\nabla^{'}_{i}
\end{eqnarray}
Here $S^{+}_{0}$ is the local positive chirality spinor bundle over $X$,
$L^{\frac{1}{2}}$ is the
square root of the line bundle $L$ over $X$,
$\nabla_{i}=\partial_{i}+\Gamma_{i}\:$, $\Gamma_{i}$ is the induced Levi-Civita
connection on $S^{+}_{0}$. $\nabla^{'}_{i}=\partial_{i}+ia_{i}\;,$
    $a_{i}$ is the connection on $L^{\frac{1}{2}}$ over $X$.
(For the later use, notice that when the covariant derivative $\nabla_{i} =
 \partial_{i} + \Gamma_{i}$
acts on the $\gamma^{,}$s or tensors on the Riemannian manifold $X$,
 $\Gamma_{i}$ is the Levi-Civita connection.)

 The second Seiberg-Witten equation is
\begin{equation}
 F^{+}_{ij}=\langle M, S_{ij}M\rangle
\end{equation}
Here $\langle ,\rangle$ represents Hermite inner product, it is the pointwise
inner product.
      $S_{ij}=\frac{i}{4}[\gamma_i,\gamma_j]$ is the spin operater on $X$, and
      $F^+_{ij}$ is the self dual part of the curvature on the line bundle
over $X$.  However, Witten did not tell us where the second equation comes
from. Obviously, a profound understanding of the Seiberg-Witten equations must
help to promote both the applications to physics and further generalizations.

  In this paper, we shall generalize Campolattaro's viewpoint and set
up the correspondence between the spinor and the curvature on the line bundle
over the Spin$^c_4$ manifold.
We also set up the correspondence between the spinor and the $2$--differential form
on the Spin$^{c}_{4}$ manifold.
Seiberg-Witten equations and the perturbed Seiberg-Witten equations can then
be well understood from this ``direct'' physical point of view.

We further suggest that the second Seiberg-Witten equation (18) (also the second
perturbed Seiberg-Witten equation) can be considered as a nonlinear
Dirac-like equation.

Since the connection is more basic than the curvature, we shall
derive the spinor representation of the connection  on the associated unitary
line bundle on the Spin$^c_4$ manifold.

\section{SPINOR REPRESENTATION OF THE CURVATURE ON THE LINE BUNDLE OVER
$Spin^{c}_{4}$ MANIFOLD}

\subsection{THE CASE OF CURVED SPACE--TIME }

In order to obtain the spinor representation of the curvature of a unitary
connection on
the line bundle over Spin$^{c}_{4}$ manifold, we first consider the case when
the space--time $X$ is a manifold with a pseudo-Riemannian metric of Lorentz
signature. As described in more detail in
(Misner, Thorne and Wheeler,1973), the Maxwell equations in the curved
space--time are
\begin{eqnarray}
 F_{\mu\nu, \gamma} + F_{\nu\gamma, \mu} + F_{\gamma\mu, \nu} &=& 0        \\
        F^{\mu\nu}_{\;\ \, ;\nu} &=& j^{\mu}
\end{eqnarray}
Where $\mu, \nu, \gamma =0, 1, 2, 3.$  These equations have the same form
as in the case of Minkowski space--time, but with a comma replaced by a
semicolon in Eq.(20). Here the semicolon stands for the covariant derivative.

One can naturally assume that there exists a spinor $\Psi$ on the curved
space--time $X$, such that
\begin{eqnarray}
F^{\mu\nu}&=&\overline{\Psi} S^{\mu\nu}\Psi           \\
\ast F^{\mu\nu}&=&\overline{\Psi} \gamma^{5} S^{\mu\nu}\Psi
\end{eqnarray}
where $\mu, \nu =0, 1, 2, 3.$
Notice that at this time $\Psi$ and $\gamma^{\mu}$ is defined on the curved
space-time.

We can now ask the question: When is $iF$ the curvature of a
unitary connction on some line bundle over the curved space-time $X$ with the
Hermitian metric?

One necessary condition is that $F$ satisfy the Bianchi identity $dF=0$,
This is just one of the Maxwell equations (19). The other is the first Chern
class $c_{1}(L)$ of a line bundle $L$ over $X$ is ``quantized" -- $c_{1}(L)$
integrates to an integer over any two dimensional cycle in $X$.
This fact can be interpreted as requiring quantization of electric charge.
Maintaining a perfect duality between electric and magnetic fields would then
require quantization of magnetic charge as well.

From now on, we suppose that both electric and  magnetic charges are
quantized. The second condition is then automatically satisfied.
We claim that these two conditions are also sufficient.

Now Eq.(21) are automatically the spinor representation of the curvature on
the line bundle $L$ over the curved space--time $X$.

\subsection{THE CASE OF $Spin^{c}_{4}$ MANIFOLD}

Let $X$ be an oriented, closed four-dimensional Riemannian manifold.
A Spin$^{c}$ structure exists on any oriented four-manifold (Hirzebruch
and Hopf, 1958; Lawson and Michelson, 1989).

The curvature $F$ of a unitary connection on the line bundle $L$ over
Spin$^{c}_{4}$-manifold $X$ satisfies the Bianchi identity,
\begin{equation}
               d F = 0
\end{equation}
We also denote
\begin{equation}
    \nabla_{j}  F_{ij} = j_{i}
\end{equation}
Here the covariant derivative $\nabla_{j} = \partial_{j} + \Gamma_{j}$,
$\Gamma_{j}$ is the Levi-Civita connection on $X$.

From the previous discussions, we have naturally the spinor representation of
the
curvature $F$,
\begin{eqnarray}
F_{i j}&=&W^{\dagger} S_{i j} W      \nonumber     \\
          &=&\langle W, S_{i j} W \rangle    \\
\ast F_{i j}&=&\langle W, \gamma_{5} S_{i j} W \rangle
\end{eqnarray}
Here $W$ is a spinor on $X$, $i, j =1,2,3,4$.

\section{SPINOR REPRESENTATION OF $2$--FORMS ON THE $Spin^{c}_{4}$ MANIFOLD}

\subsection{THE CASE OF MINKOWSKI SPACE--TIME}

Campolattaro (1990a, 1990b) assumed that together with an
electric current $j_{\mu}$, there exists also a magnetic monopole current
$g_{\mu}.$ Maxwell equations read
\begin{eqnarray}
          (\ast F'^{\mu\nu})_{,\nu} &=& g^{\mu}           \\
        F'^{\mu\nu}_{\;\;\;\; ,\nu} &=& j^{\mu}
\end{eqnarray}
There exists a spinor, such that
\begin{eqnarray}
F'^{\mu\nu}&=&\overline{\Psi} S^{\mu\nu}\Psi           \\
\ast F'^{\mu\nu}&=&\overline{\Psi} \gamma^{5} S^{\mu\nu}\Psi
\end{eqnarray}
It was shown that the spinor equation Eq.(13) in the presence of magnetic
monopoles, reads
\begin{equation}
 \gamma^{\mu}\Psi_{,\mu} = -i\gamma^{\mu}\frac{e^{\gamma^{5}\alpha}}{\rho}
  \{ Im(\overline{\Psi}_{,\mu}\Psi)- j_{\mu}-\gamma^{5}
  [Im(\overline{\Psi}_{,\mu} \gamma^{5}\Psi) -g_{\mu}] \} \Psi
\end{equation}

\subsection{THE CASE OF CURVED SPACE--TIME}

Given $X$, a four-dimensional manifold with a pseudo-Rimaniann metric of
Lorentz signature. The generalized Maxwell equations read
\begin{eqnarray}
          (\ast F'^{\mu\nu})_{;\nu} &=& g^{\mu}           \\
        F'^{\mu\nu}_{\;\;\;\; ;\nu} &=& j^{\mu}
\end{eqnarray}
One has the same expressions as in the previous discussions,
\begin{eqnarray}
F'^{\mu\nu}&=&\overline{\Psi} S^{\mu\nu}\Psi           \\
\ast F'^{\mu\nu}&=&\overline{\Psi} \gamma^{5} S^{\mu\nu}\Psi
\end{eqnarray}
Notice that at this time $\gamma^{\mu}$ and $\Psi$ are defined in the curved
space--time.

From Eq.(32), we state that the Bianchi identity is no longer satisfied:
$F'$ is only a $2$--differential form on $X$. Given a $2$-form $F'$ one has
\begin{equation}
               F'= F + \omega
\end{equation}
Here $F$ is the curvature of a unitary connection on some line bundle over
 $X$, $\omega$ is a $2$-form over $X$. Later we shall adopt this notation.

\subsection{THE CASE OF $Spin^{c}_{4}$ MANIFOLD}

Denote $X$ a Spin$^{c}_{4}$ manifold. Given a $2$-form
$F+\omega$
on $X$. Denote
\begin{eqnarray}
   \nabla_{j} (\ast F_{ij} + \ast \omega_{ij})  &=& g_{i}              \\
   \nabla_{j} ( F_{ij} +  \omega_{ij})  &=& j_{i}
\end{eqnarray}
We have natually the spinor representation
\begin{eqnarray}
F_{i j} +\omega_{i j}    &=&\langle W, S_{i j} W \rangle    \\
\ast F_{i j} +\ast\omega_{i j} &=&\langle W, \gamma_{5} S_{i j} W \rangle
\end{eqnarray}
Here $W$ is a spinor on $X$, $i, j =1,2,3,4$.

\section{NONLINEAR DIRAC-LIKE EQUATION AND THE SECOND SEIBERG-WITTEN EQUATION}

From Eqs.(25) and (26),
  The self-dual part of the curvature on $L$ over the Spin$^{c}_{4}$ manifold
 $X$ is
\begin{eqnarray}
  F^{+}_{ij}&=&\frac{1}{2}(F_{ij}+ *F_{ij})\nonumber\\
            &=&\langle W, \frac{1+\gamma_{5}}{2} S_{ij} W\rangle
\end{eqnarray}
Denote $\frac{1+\gamma_{5}}{2} W =  M$, Then
\begin{equation}
   F^{+}_{ij}=\langle M, S_{ij} M\rangle      \label{X}
\end{equation}
 Eq.(\ref{X}) is just the second Seiberg-Witten equation (18).
From Eqs.(23) and (24), one has
\begin{equation}
\nabla_{i} F^{+}_{ji} = \nabla_{i}\langle M, S_{ji} M \rangle
                      =\frac{1}{2}j_{j}
\end{equation}
Notice that
$\widetilde{\nabla_{i}}S_{ji}
                  =\nabla_{i}S_{ji}$. Eq.(42) has the
equivalent form
\begin{equation}
  Im\langle M, \gamma_{j} \widetilde{D} M\rangle=
\langle M, \nabla_{i}S_{ji}M\rangle
-Im\langle M, \widetilde{\nabla_{j}}M\rangle-\frac{1}{2}j_{j}
\end{equation}
Just as Campolattaro's considerations, given $F^{+}_{ij}$ thus $j_{j}$, one
can verify that the positive chirality spinor $M$ satisfy the nonlinear
Dirac-like equation on the Spin$^{c}_{4}$ manifold $X$
\begin{equation}
 \widetilde{D} M= -\frac{i}{\langle M, M \rangle }
\{\langle M, \nabla_{k}S_{ik} M\rangle
-Im\langle M, \widetilde{\nabla_{i}}M\rangle-\frac{1}{2}j_{i}\} \gamma_{i} M
\end{equation}
Since the term $\langle M, \gamma_{j}\gamma_{i} M \rangle$ is  pure
imaginary, if $i \ne j$. One can show that Eq.(45)  is the sufficient
condition of Eq.(44), thus is the sufficient
condition of the second Seiberg-Witten equation (42).

Notice that Seiberg-Witten equations
\begin{eqnarray*}
 \widetilde{D} M &=& 0               \\
    F^{+}_{ij} &=& \langle M, S_{i j} M \rangle
\end{eqnarray*}
are equivalent to
\begin{eqnarray*}
 \widetilde{D} M &=& 0                 \\
 \widetilde{D} M &=& -\frac{i}{\langle M, M \rangle }
\{\langle M, \nabla_{k}S_{ik} M\rangle
-Im\langle M, \widetilde{\nabla_{i}}M\rangle-\frac{1}{2}j_{i}\}
\gamma_{i} M
\end{eqnarray*}
From this point of view, the second Seiberg-Witten equation can be replaced
by the nonlinear Dirac-like equation (45).

\section{NONLINEAR DIRAC-LIKE EQUATION AND
        THE PERTURBED SECOND SEIBERG-WITTEN EQUATION}

From Eqs.(39) and (40), write $ \phi = \omega^{+} = \frac{1}{2}(\omega
+ \ast\omega)$, one has
\begin{equation}
   F^{+}_{ij} + \phi_{i j}=\langle M, S_{ij} M\rangle      \label{Y}
\end{equation}
 Eq.(\ref{Y}) is just the second perturbed Seiberg-Witten equation.
From Eqs.(37) and (38), we have
\begin{equation}
\nabla_{i}\langle M, S_{ji} M \rangle =\frac{j_{j}+g_{j}}{2}
\end{equation}
Just as the previous discussions, given $j_{j}+ g_{j}$, we have a nonlinear
Dirac-like equation on the Spin$^{c}_{4}$ manifold $X$,
\begin{equation}
 \widetilde{D} M= -\frac{i}{\langle M, M \rangle }
\{\langle M, \nabla_{k}S_{ik}M\rangle
-Im\langle M, \widetilde{\nabla_{i}}M\rangle-\frac{j_{i}+g_{i}}{2}\}
\gamma_{i} M
\end{equation}
This equation is the sufficient condition of the second perturbed
Seiberg-Witten equation (\ref{Y}).
Notice that the perturbed Seiberg-Witten equations
\begin{eqnarray*}
 \widetilde{D} M &=& 0               \\
    F^{+}_{ij}  + \phi_{ij}&=& \langle M, S_{i j} M \rangle
\end{eqnarray*}
are equivalent to
\begin{eqnarray*}
 \widetilde{D} M &=& 0  \\
 \widetilde{D} M &=& -\frac{i}{\langle M, M \rangle }
\{\langle M, \nabla_{k}S_{ik}M\rangle
-Im\langle M, \widetilde{\nabla_{i}}M\rangle-\frac{j_{i}+g_{i}}{2}\}
\gamma_{i} M
\end{eqnarray*}
So one can say that it is important to study the nonlinear Dirac-like
equations (45) and (48).

\section{SPINOR REPRESENTATION OF  $U(1)$ CONNECTION ON
$Spin^{c}_{4}$ MANIFOLD}

 Now we derive the spinor representation of the connection $A$ on the line
 bundle $L$ over Spin$^{c}_{4}$ manifold $X$. Choose a positive chirality
spinor $M$, which is the solution of the Dirac equation
\begin{equation}
   \gamma_{i}\widetilde{\nabla_{i}} M =0         \label{E}
\end{equation}
One has
\begin {equation}
 \langle M,\gamma_{j}\gamma_{i}\widetilde{\nabla_{i}} M\rangle=0
\end{equation}
which is equal to
\begin{equation}
      \langle M, S_{ji}\widetilde{\nabla_{i}}M\rangle
     -\frac{i}{2}\langle M,\widetilde{\nabla_{j}}M\rangle=0    \label{H}
\end{equation}
By taking the Hermitian conjugation of Eq.(\ref{H}), one has
\begin{equation}
      \langle\widetilde{\nabla_{i}}M, S_{ji} M\rangle
     +\frac{i}{2}\langle\widetilde{\nabla_{j}}M, M\rangle=0     \label{I}
\end{equation}
Notice that $\widetilde{\nabla_{i}}S_{ji}=\nabla_{i}S_{ji}\:$.
By adding Eqs.(\ref{H}) and (\ref{I}), one obtains,
\begin{equation}
   \nabla_{i}\langle M, S_{ji}  M\rangle
                     =\langle M,(\nabla_{i}S_{ji}) M\rangle
                       -Im\langle M,\widetilde{\nabla_{j}}M\rangle   \label{M}
\end{equation}
Eq.(\ref{M}) is completely equivalent to the Dirac Eq.(\ref{E}).

  Since the computation is local, we can write
\begin{equation}
         M  =  \Psi\otimes \lambda
\end{equation}
where $\Psi\in S^{+}_{0}$ is a local positive chirality spinor
 on $X$, and $\lambda\in L^{\frac{1}{2}}$.
Notice that
\begin{equation}
   \langle\Psi_{1}\otimes\lambda_{1},\Psi_{2}\otimes\lambda_{2}\rangle
  =\langle\Psi_{1},\Psi_{2}\rangle\langle\lambda_{1},\lambda_{2}\rangle
\end{equation}
Eq.(\ref{M}) reads
\begin{eqnarray}
  \nabla_{i}[\langle\Psi, S_{ji}\Psi\rangle
               \langle\lambda,\lambda\rangle]
&=&\langle\Psi,(\nabla_{i}S_{ji})\Psi\rangle
  \langle\lambda,\lambda\rangle          \nonumber   \\
 & &\mbox{}-Im\langle\Psi\otimes\lambda,\nabla_{j}\Psi\otimes\lambda
                  +\Psi\otimes\nabla^{'}_{j}\lambda\rangle      \label{LZ}
\end{eqnarray}
Here $\nabla_{j}=\partial_{j}+\Gamma_{j}\:$.
     $\nabla^{'}_{j}=\partial_{j}+ia_{j}$, $a_{j}$ is the connection on
$L^{\frac{1}{2}}\:$.
We have
\begin{eqnarray}
a_{j}&=&  -\frac{Im\langle\Psi,\nabla_{j}\Psi\rangle}
                 {\langle\Psi,\Psi\rangle}
         +\frac{\langle\Psi,(\nabla_{i}S_{ji})\Psi\rangle}
               {\langle\Psi,\Psi\rangle}               \nonumber        \\
      & &\mbox{}- \frac{\nabla_{i}[\langle\Psi, S_{ji}\Psi\rangle
                                  \langle\lambda,\lambda\rangle]}
               {\langle\Psi,\Psi\rangle\langle\lambda,\lambda\rangle}
         -\frac{Im\langle\lambda,\partial_{j}\lambda\rangle}
                {\langle\lambda,\lambda\rangle}
\end{eqnarray}
Since $L$ is unitary, i.e.,
      $\langle\lambda,\lambda\rangle=1.\ \ $ One has
\[  Im\langle\lambda,\partial_{j}\lambda\rangle
             =-i\lambda^{-1}\partial_{j}\lambda \]
The transformation rule of $a_{j}$ is
\begin{equation}
   ia'_{j}=ia_{j}+\lambda^{-1}\partial_{j}\lambda
\end{equation}
We also denote $a'$ by $a$.
It leads to
\begin{eqnarray}
    a_{j} &=& -\frac{Im\langle\Psi,\nabla_{j}\Psi\rangle}
                       {\langle\Psi,\Psi\rangle}
               +\frac{\langle\Psi,(\nabla_{i}S_{ji})\Psi\rangle}
                     {\langle\Psi,\Psi\rangle}
               -\frac{\nabla_{i}\langle\Psi, S_{ji}\Psi\rangle}
                     {\langle\Psi,\Psi\rangle}     \nonumber  \\
  &=& \frac{Im\langle\Psi, \gamma_{j} D \Psi \rangle}
                       {\langle\Psi,\Psi\rangle}.
\end{eqnarray}
Here $D = \gamma_{i} \nabla_{i}$ is the Dirac operator acting on the
 local positive chirality spinor $\Psi$.  This means that the Dirac equation
 (49) can be locally reduced to
\begin{equation}
   \gamma_{j}(\nabla_{j} + ia_{j}) \Psi =0
\end{equation}
Notice that $da=\frac{1}{2}F$. For any unitary connection $A$ on $L$,
$F=dA$. we can choose
\begin{equation}
           A=2a
\end{equation}
Finally, one has the spinor representation of $A$,
\begin{equation}
    A_{j} = 2 \frac{Im\langle\Psi, \gamma_{j} D \Psi \rangle}
                       {\langle\Psi,\Psi\rangle}.
\end{equation}
Where $j=1, 2, 3, 4.$

\section{CONCLUSIONS}

We have analysed the relation between the
spinor and the curvature on a unitary line bundle over the Spin$^c_4$
manifold. The relation between the spinor and the $2$--form on the
Spin$^{c}_{4}$ manifold has also been  considered.

We have proposed two nonlinear Dirac-like equations which are equivalent
to the second Seiberg-Witten equation and the second perturbed Seiberg-Witten
equation respectively. This means one can study (the self dual part of) the
curvature (or the $2$-form ) in the quantum mechanics level.

We have also derived the local spinor representation of the unitary connection
 on the associated line
bundle over the Spin$^c_4$ manifold.

These discussions can be generalized to the non-Abelian cases.

The relation between the properties of the nonlinear Dirac-like equation and
the moduli space of the (perturbed) Seiberg-Witten equations need to be
further studied.

{\Large \bf ACKNOWLEDGMENTS}

 One of us (L. Z. Hu) woud like to thank Profs. M.-L. Ge and Y. S. Wu
 for valuable discussions.
  This work is supported in part by the Research Center of Mathematics,
  the National  Education Committee of China.

\end{document}